\documentclass[10pt,journal]{IEEEtran}

\ifCLASSINFOpdf

\else

\fi

\usepackage{mathrsfs}
\usepackage{amsmath}
\usepackage{amsfonts}
\usepackage{amsthm}
\usepackage{amssymb}
\usepackage{graphicx}
\usepackage{subfigure}
\usepackage{indentfirst}
\usepackage{array}
\usepackage{epstopdf}
\usepackage{cite}
\usepackage{enumerate}
\usepackage{bm}
\usepackage{dsfont}
\usepackage{multirow}
\usepackage{verbatim}
\usepackage{stfloats}
\usepackage{placeins}
\usepackage{makecell}
\usepackage{cancel}
\usepackage{threeparttable}
\usepackage{tikz}
\usetikzlibrary{positioning}
\usetikzlibrary{spy}
\usepackage{wrapfig}
\usepackage{pgfplots}
\usepackage{nicefrac}
\usepackage{thmtools,thm-restate}
\usepackage{color}

\usepackage{booktabs}
\usepackage{graphicx}
\usepackage{bbm}
\usepackage{adjustbox}
\usepackage{multirow}
\usepackage{tabularx}
\usepackage[misc]{ifsym}
\usepackage{bbding}
\usepackage{colortbl}

\newcolumntype{Y}{>{\centering\arraybackslash}X}
\newcolumntype{C}[1]{>{\centering\arraybackslash}m{#1}}
\usepackage{algorithm}
\usepackage{algorithmicx}
\usepackage{algpseudocode}
\usepackage{xcolor}
\usepackage[linkcolor=blue,citecolor=blue,colorlinks=true,urlcolor=blue]{hyperref}

\algnewcommand{\LineComment}[1]{\State {\color{gray}\(\triangleright\) #1}}
\definecolor{lightblue}{rgb}{0.9, 0.92, 0.96}
\definecolor{subsectioncolor}{rgb}{0, 0, 0.706}   
\definecolor{mylightblue}{rgb}{0.851, 0.898, 0.941} %
\definecolor{LightCyan}{rgb}{0.88,1,1}
\definecolor{LightOrange}{rgb}{1,0.85,0.70}

\usepackage{pifont}
\definecolor{C0}{rgb}{0.121569, 0.466667, 0.705882}
\definecolor{C1}{rgb}{1.000000, 0.498039, 0.054902}
\definecolor{C2}{rgb}{0.172549, 0.627451, 0.172549}
\definecolor{C3}{rgb}{0.839216, 0.152941, 0.156863}
\definecolor{C4}{rgb}{0.580392, 0.403922, 0.741176}
\definecolor{C5}{rgb}{0.549020, 0.337255, 0.294118}
\definecolor{C6}{rgb}{0.890196, 0.466667, 0.760784}
\definecolor{C7}{rgb}{0.498039, 0.498039, 0.498039}
\definecolor{C8}{rgb}{0.737255, 0.741176, 0.133333}
\definecolor{C9}{rgb}{0.090196, 0.745098, 0.811765}
\definecolor{trolleygrey}{rgb}{0.5, 0.5, 0.5}

\usepackage{enumitem}

\begin{document}
	
	\title{Tokenized Joint Source-Network Coding for Loss-Resilient Video Streaming}

	\author{Ke Yang,
		Sixian Wang,~\IEEEmembership{Member,~IEEE},
		Kailin Tan,~\IEEEmembership{Graduate Student Member,~IEEE},
		Guo Lu,~\IEEEmembership{Member,~IEEE},
		Jincheng Dai,~\IEEEmembership{Member, IEEE},
		Wenjun Zhang,~\IEEEmembership{Fellow, IEEE},
		and Ping Zhang,~\IEEEmembership{Fellow, IEEE}
		
		\thanks{This work was supported in part by the National Key Research and Development Program of China under Grant 2024YFF0509700, in part by the National Natural Science Foundation of China under Grant 62371063, Grant 62601519, and Grant 62471290, in part by the Beijing Municipal Natural Science Foundation under Grant L232047, in part by Postdoctoral Fellowship Program of CPSF under Grant Number GZB20250810, in part by the China Postdoctoral Science Foundation under Grant Number 2025M783515, and in part by the Beijing Nova Program. \emph{(Ke Yang and Sixian Wang are co-first authors.) (Corresponding author: Jincheng Dai.)}}
		
		\thanks{Ke Yang, Kailin Tan, Jincheng Dai, and Ping Zhang are with Beijing University of Posts and Telecommunications, Beijing 100876, China. (email: daijincheng@bupt.edu.cn)}
		
		\thanks{Sixian Wang, Guo Lu, and Wenjun Zhang are with Shanghai Jiao Tong University, Shanghai 200240, China.}
		
		\vspace{-1em}
	}
	
	\maketitle

	\begin{abstract}
		For decades, conventional video streaming systems have relied on packet-level forward error correction (FEC) codes and retransmissions to ensure reliable video delivery. However, in real-time communications, FEC is inherently inefficient under unknown and time-varying packet-loss conditions, while retransmission can incur excessive latency and fail to meet stringent playout deadlines. Emerging deep joint source-channel coding (DeepJSCC) provides an alternative by learning channel-noise-resilient representations, yet its tight coupling with the physical layer makes it difficult to integrate into today’s layered network architectures and protocols. In this paper, we propose TokJSNC, a simple-but-effective tokenized joint source-network coding paradigm for packet-loss-resilient video streaming. Rather than first compressing video into a compact bitstream and then protecting it with packet-level FEC, TokJSNC directly encodes video into variable-length discrete token-stream, whose transmission rate is jointly adapted to video content complexity and network conditions. By exploiting the spatiotemporal correlations among learned video tokens, TokJSNC enables robust video reconstruction over a wide range of packet-loss rates without explicit FEC or retransmissions. Consequently, TokJSNC brings the JSCC-like graceful degradation property to the network layer while remaining compatible with existing network architectures and transport protocols. Experimental results show that TokJSNC maintains an average playback frame rate above 22 fps even at a 70\% packet-loss rate, whereas the best-performing baseline drops below 12 fps. Overall, TokJSNC establishes a practical pathway for repurposing modern generative video tokenizers for loss-resilient streaming, while providing a fundamental framework for video token communications.
	\end{abstract}

	
	\IEEEpeerreviewmaketitle
	
	\section{Introduction}
	
	Real-time communications (RTC) have become essential for applications including video conferencing, cloud gaming, teleoperation, mobile surveillance, and immersive interaction. 
	Unlike video-on-demand services, RTC applications typically operate under strict latency requirements \cite{fouladi2018salsify}.
	When streaming over degraded network conditions, packets may be corrupted or arrive after their playout deadlines, while retransmission requests may not complete in time \cite{cheng2024grace}.
	In this case, an RTC receiver may have to reconstruct video from partially received packets, which poses significant challenges on preserving video quality and continuity.
	
	\begin{figure}[t]
		\centering
		\includegraphics[width=0.49\textwidth]{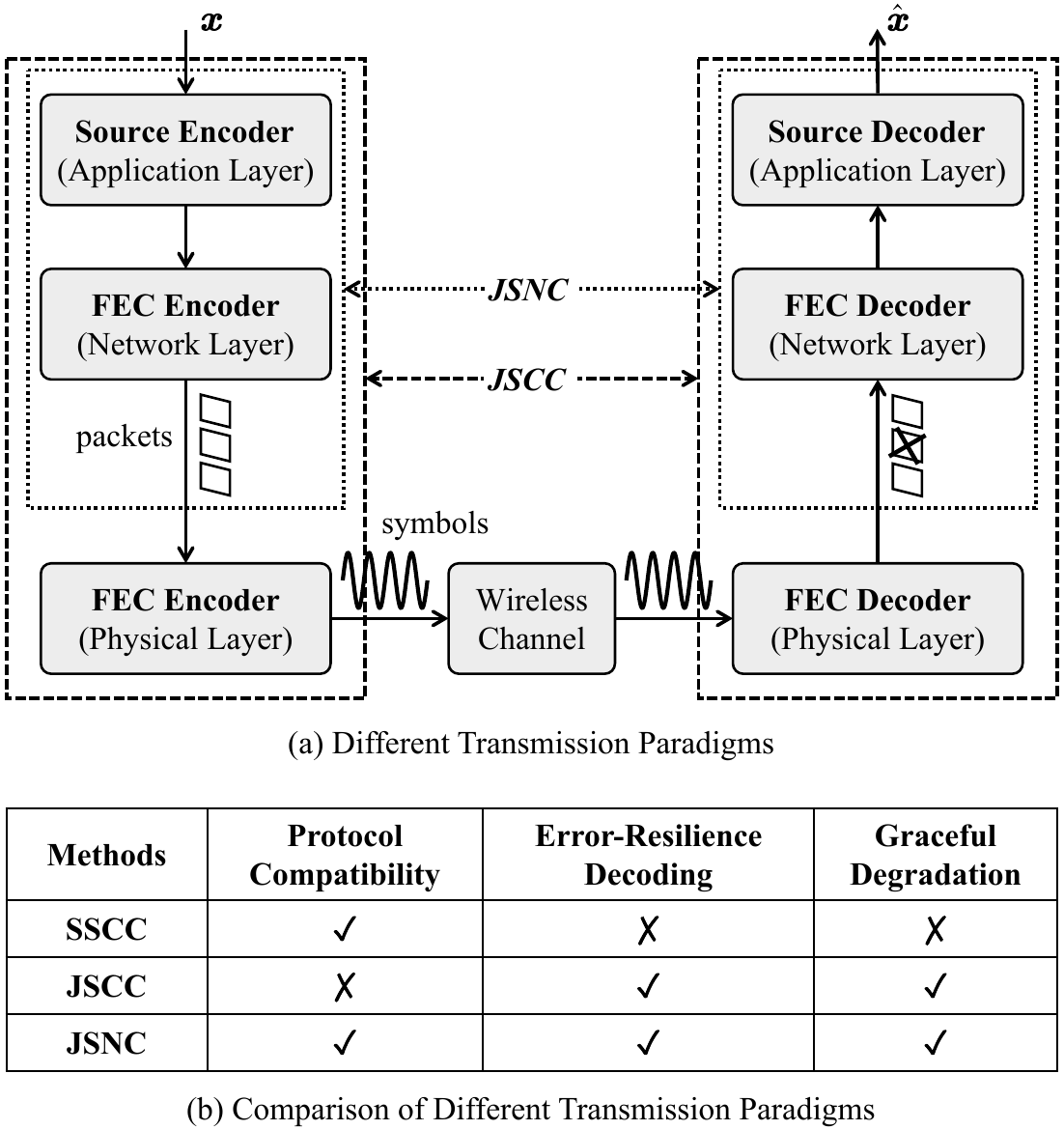}
		\vspace{-0.5cm}
		\caption{Comparison of different transmission paradigms. (a) Conceptual illustration of the interaction between source coding and network delivery. SSCC relies on independent source compression and network-layer protection, JSCC jointly optimizes source and channel coding but sacrifices compatibility with existing protocols, while JSNC introduces cross-layer coordination to achieve both protocol compatibility and error-resilient transmission. (b) Qualitative comparison of SSCC, JSCC, and JSNC regarding protocol compatibility, error-resilient decoding, and graceful degradation.}
		\label{fig:paragram_comparsion}
		\vspace{-1em}
	\end{figure}
	
	Existing video streaming systems typically adopt a multi-stage architecture, where source compression, packetized network transport, and loss protection are designed and operated at different layers independently.
	For the source compression stage, modern hybrid video codecs, such as H.264/AVC \cite{wiegand2003overview} and H.265/HEVC \cite{sze2014high}, achieve strong rate-distortion performance by compressing video into compact bitstreams, while neural video codecs, including DVC \cite{lu2019dvc} and the DCVC series \cite{li2024neural,jia2025towards}, further improve compression efficiency through learned prediction, context, and entropy models.
	Such compact representations typically contain strong spatial and temporal dependencies, such as syntax elements, prediction references, or entropy-decoding contexts.
	When transmission errors corrupt or erase such dependency information, the decoder may lose synchronization with the encoder, resulting in decoding failures and severe temporal error propagation that can persist until the next intra-coded refresh.
	To protect the fragile bitstream, current video streaming systems commonly introduce retransmission or forward error correction (FEC). 
	Retransmitting lost packets is often infeasible under stringent latency requirements, while FEC introduces additional bandwidth overhead and must add redundancy before packet losses occur \cite{cheng2024grace}.
	Under stringent deadlines and dynamically varying losses, this separation makes it difficult to achieve both high compression efficiency and loss-resilient reconstruction.

	This limitation suggests that loss resilience should be incorporated into the construction of the source representation itself, rather than being added only after compression. 
	Following this principle, recent joint source-channel coding (JSCC) \cite{DJSCC, yang2024swinjscc} studies directly map pixel values to the complex-valued symbols for physical-channel transmission, which exhibit graceful performance degradation as the channel SNR varies.
	However, such a graceful property relies on end-to-end optimization from source compression in the application layer to channel coding and modulation in the physical layer, which requires significant changes to integrate with today's packet-based layered network infrastructures \cite{gunduz2025joint}.
	How to achieve graceful performance degradation while retain compatibility to current network architectures?
	Motivated by DeepJSCC \cite{DJSCC}, we propose a joint source-network coding (JSNC) perspective, which extends the joint-design principle to the source representation and packet-loss process.
	In JSNC, the source representation is jointly optimized with packetization and packet loss concealment, and the packet lossy network is considered as a non-trainable layer in the middle.
	Fig.~\ref{fig:paragram_comparsion} compares the separated source and channel coding (SSCC), JSCC, and JSNC paradigms.
	
	To build such a JSNC framework, a key requirement is that the source representation remain usable when only a subset of transmitted packets is successfully received. To this end, we draw inspiration from recent advances in visual tokenization \cite{yu2023magvit,yu2024language,luo2024open,NEURIPS2024_31994923}. Modern video tokenizers \cite{nvidia2025cosmos} map high-dimensional video signals into compact spatiotemporal tokens, providing modular representation units that can be naturally organized for packetized transmission. More importantly, lookup-free quantization (LFQ) \cite{yu2024language} quantizes latent features into discrete binary-valued tokens without requiring an internal entropy coder. 
	As a result, when some transmitted tokens are lost, the remaining tokens can still be independently dequantized. 
	In this way, the impact of packet loss is equated to the loss of token groups. 
	Such a property makes discrete visual tokens a suitable interface for JSNC, making reconstruction from partial packets feasible. 
	
	Nevertheless, visual tokens alone do not guarantee efficient and stable video streaming over packet-loss networks. 
	Three system-level challenges remain. 
	First, token allocation should adapt to the information density of video content.
	Recent adaptive tokenizers show that complex videos benefit from larger token budgets, while simple or predictable content can be represented with fewer tokens \cite{ye2026infotok}.
	Thus both how many tokens to transmit and which tokens contribute most to reconstruction should be modeled.
	Second, token packetization should disperse neighboring spatiotemporal tokens across different packets, so that packet losses result in distributed missing observations rather than large contiguous missing regions that are difficult to reconstruct.
	Third, token encoding and reconstruction should adapt to the current network state rather than assuming a fixed packet-loss rate. The encoder can condition the transmitted representation on recent packet-loss statistics, while the decoder can further exploit the observed token-availability pattern to recover the missing representation.
	
	To address these issues, we propose TokJSNC, a tokenized joint source-network coding framework for video streaming over packet-loss networks. 
	TokJSNC retains discrete token quantization as the transmission interface, such that successfully received transport tokens remain usable when other packets are lost. 
	A reconstruction-proxy-guided token router performs content-adaptive allocation at two levels: clip-level reconstruction complexity determines the token budget, while localized reconstruction errors guide the selection of informative spatiotemporal positions. The selected tokens are further organized through spatiotemporal token packetization to reduce the concentration of missing observations after packet erasures. In addition, a network-state-conditioned TokJSNC encoder adapts the transport representation to slow-timescale packet-loss statistics, while a state-aware decoder combines the shared network state prior with the observed token states to recover incomplete latent-token representations. In this way, TokJSNC jointly adapts token allocation, packet organization, transport representation, and latent-token recovery without requiring knowledge of future packet-loss realizations.
	
	The main contributions of this paper are summarized as follows:
	\begin{itemize}
		\item We propose TokJSNC, a tokenized joint source-network coding framework that enables reconstruction from partial packets and retains compatibility with current layered network architectures.
		
		\item We develop a reconstruction-proxy-guided adaptive token routing mechanism that determines the clip-level token budget according to reconstruction-derived content complexity and prioritizes informative spatiotemporal positions within the available budget.
		
		\item We design network-state-conditioned token encoding and state-aware decoding that exploit shared packet-loss statistics and observed token states. Together with spatiotemporal token packetization, the proposed design enables graceful recovery from incomplete token representations without relying on FEC or retransmission.
		
	\end{itemize}
	
	The remainder of this paper is organized as follows. Section \ref{related_work} reviews related work on conventional and neural video coding, neural video tokenization, and loss-resilient neural video coding. Section \ref{method} presents the proposed TokJSNC framework. Section \ref{experiments} reports the experimental results and ablation studies. Section \ref{conclusion} concludes this paper.

	\section{Related Work} \label{related_work}
	
	\subsection{Conventional and Neural Video Coding}
	
	Traditional video codecs, including H.264/AVC~\cite{wiegand2003overview}, H.265/HEVC~\cite{sze2014high}, and H.266/VVC~\cite{bross2021developments}, exploit spatial and temporal redundancies through predictive and transform-based coding to achieve high compression efficiency. However, their compressed bitstreams contain strong dependencies among syntax elements, motion information, residuals, and reference frames. Packet loss can therefore cause decoding failures or error propagation beyond the corrupted payload, making reliable delivery essential for real-time video streaming.
	
	Neural video codecs further improve compression efficiency by learning compact latent representations and exploiting learned prediction and context models. Representative works including DVC~\cite{lu2019dvc} and the DCVC series~\cite{li2021deep,li2022hybrid,li2023neural,li2024neural,jia2025towards} progressively improve compression efficiency through learned motion, spatiotemporal context, and feature representations. However, these codecs similarly rely on reliable delivery of their compressed representations and generally do not explicitly optimize reconstruction from incomplete packet reception. Consequently, packet losses can still lead to decoding failures or severe reconstruction degradation.
	
	\subsection{Neural Video Tokenization}
	
	Recent advances in generative video models have established neural video tokenization as an effective approach for compact visual representation. MAGVIT~\cite{yu2023magvit} and Open-MAGVIT2~\cite{luo2024open} learn discrete spatiotemporal token representations for efficient video generation and reconstruction, while Cosmos Tokenizer~\cite{nvidia2025cosmos} further demonstrates the scalability of visual tokens as a general-purpose representation. LFQ~\cite{yu2024language} provides a codebook-free approach for discrete visual tokenization, enabling token representations without explicit learned codebooks. OmniTokenizer~\cite{NEURIPS2024_31994923} further explores unified tokenization across image and video modalities.
	
	Compared with conventional compressed bitstreams, discrete token representations provide a natural basis for partial packet reception because individual tokens can remain available when other tokens are lost. However, existing video tokenizers are primarily designed for generation or representation learning rather than packetized communication. Their token granularity and representation allocation are generally not adapted to heterogeneous network conditions, leaving content-adaptive token allocation, packet organization, and network-aware recovery insufficiently explored.
	
	\subsection{Loss-Resilient Neural Video Coding} 
	
	Recent studies have incorporated packet-loss resilience into learned visual compression, enabling reconstruction from incomplete packet reception. GRACE~\cite{cheng2024grace,cheng2023graceplusplus} enables real-time video reconstruction under packet loss through random-mask training and packet-level entropy coding, demonstrating graceful quality degradation as fewer packets are received. ResiComp~\cite{wang2025resicomp}, in contrast, is a loss-resilient image compression method that recovers missing latent representations through masked visual token modeling, without addressing the temporal dependencies and playback-continuity requirements of video streaming. These methods establish the feasibility of reconstructing learned representations from partial packet reception, but mainly focus on improving the recoverability of a given transmitted representation. They do not explicitly formulate content-adaptive source allocation, network-state-conditioned representation, spatiotemporal packet organization, and receiver-side recovery as a unified source-network coding problem. Building on these advances, TokJSNC introduces the JSNC perspective and jointly coordinates these components according to both video content and packet-loss conditions.
	
	\section{Method}\label{method}
	
	\subsection{Overview}
	
\begin{figure*}[t]
		\centering
		\includegraphics[width=0.99\textwidth]{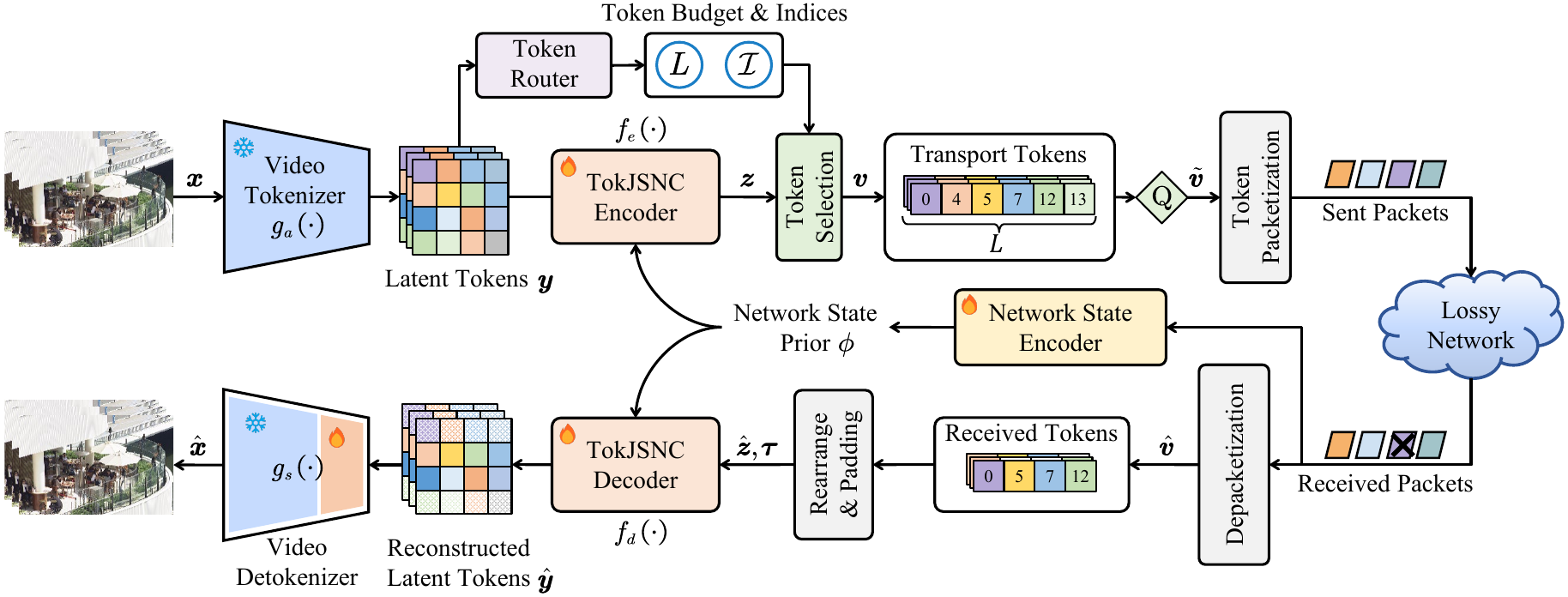}
		\caption{Overall architecture of TokJSNC for packet-loss-resilient video streaming. }
		\label{fig:framework}
	\end{figure*}

	TokJSNC is a token-level joint source-network coding framework for packet-loss-resilient video transmission. Unlike conventional video coding schemes that optimize source compression independently of network delivery, TokJSNC jointly considers source representation adaptation, network-aware token transmission, and latent-token recovery under packet-loss networks. As illustrated in Fig.~\ref{fig:framework}, the major components of TokJSNC are summarized as follows.

	\begin{itemize}
		
		\item \textbf{Video Tokenization}:  
		Given an input video clip $\bm{x}\in\mathbb{R}^{T\times H\times W\times3}$, the pretrained video tokenizer $g_a(\cdot)$ maps the video into a spatiotemporal latent-token grid $\bm{y}=g_a(\bm{x})\in\mathbb{R}^{t\times h\times w\times c}$, where $t$, $h$, and $w$ denote the temporal and spatial dimensions of the latent grid, respectively, and $c$ is the token embedding dimension. The adopted tokenizer independently encodes the first frame and temporally compresses the remaining frames by a factor of 4, resulting in $T=4(t-1)+1$. Spatially, the tokenizer performs an $8\times$ downsampling, yielding $h=H/8$ and $w=W/8$. Therefore, the total number of latent tokens is $N=t\times h\times w$.
		
		\item \textbf{Content-Aware Token Router}:  
		The content-aware token router first estimates the content complexity $\rho$ of each video clip based on the reconstruction distortion between the original clip $\bm{x}$ and the proxy reconstruction $\tilde{\bm{x}}$. The proxy reconstruction is generated from $\bm{y}$ by an independent frozen copy of the pretrained detokenizer. Then, the router determines the adaptive token budget $L$, and $\eta=L/N$ denotes the actual token transmission ratio. Meanwhile, the token-level content complexity is exploited to identify the corresponding token positions, producing the selected token indices $\mathcal{I}$.
		
		\item \textbf{TokJSNC Encoder}:  
		The latent tokens $\bm{y}$ are transformed into transport-oriented representations $\bm{z}=f_{e}(\bm{y};\boldsymbol{\phi}, \eta)$ by the TokJSNC encoder, 
		where $\boldsymbol{\phi}$ denotes the shared network state prior. The prior is generated by a learnable network state encoder based on observed network statistics and periodically updated to characterize the average packet-loss condition over a temporal window. By incorporating $\boldsymbol{\phi}$ into the representation transformation, the encoder generates transport tokens adapted to different packet-loss conditions.
		
		\item \textbf{Token Selection}: 
		The token budget $L$ and the selected token indices $\mathcal{I}$ are used to select the transport tokens
		$\bm{v}=\operatorname{Gather}(\bm{z},\mathcal{I})$.
		
		\item \textbf{Quantization}: 
		The selected transport tokens $\bm{v}$ are quantized into discrete tokens through a quantization function: $\widetilde{\bm{v}}=\mathrm{Q}(\bm{v})$. The inverse quantization is performed at the receiver to recover the received tokens.
		
		\item \textbf{Token Packetization}:  
		The quantized transport tokens $\widetilde{\bm{v}}$ are divided into multiple groups according to a predefined spatiotemporal token packetization strategy. Tokens belonging to the same region are grouped into one packet, resulting in the packet set $\mathcal{P}=\{P_1,\ldots,P_M\}$. 
		
		\item \textbf{Packet Lossy Network}:  
		During transmission, packet loss may occur due to network conditions. Since each packet contains a subset of transport tokens, the loss of a packet results in the unavailability of the corresponding token subset. The receiver detects packet reception states through packet headers or error detection mechanisms, and the packet loss process is modeled as:
		$		(\hat{\bm{v}},\mathbf{F})
		=
		\mathcal{W}(\widetilde{\bm{v}}),$ where $\mathbf{F}=[F_1,\ldots,F_L]\in\{0,1\}^{L}$ denotes the token-reception vector. For the $l$-th selected token, $F_l=1$ if $\widetilde{\bm v}_l$ is successfully received and $F_l=0$ otherwise.
		
		\item \textbf{TokJSNC Decoder}:  
		At the receiver, $\mathbf F$ and $\mathcal I$ are used to construct the incomplete token representation for TokJSNC Decoding. Each latent position is classified into one of three states: received, dropped, or unselected. Conditioned on the shared network state prior $\boldsymbol{\phi}$ and token-state information, the TokJSNC decoder reconstructs the complete latent tokens: $\hat{\bm{y}} = f_{d}
		(\hat{\bm{z}};\boldsymbol{\tau},\boldsymbol{\phi}),$
		where $\boldsymbol{\tau}$ denotes the token-state vector.
		
		\item \textbf{Video Reconstruction}:  
		The recovered latent tokens $\hat{\bm y}$ are decoded by the video detokenizer $g_s(\cdot)$ to obtain the reconstructed video clip, i.e., $\hat{\bm{x}}=g_s(\hat{\bm{y}}).$
		
	\end{itemize}

	\subsection{Adaptive Token Router}
	\label{subsec:adaptive_token_routing}
	
	Most existing video tokenizers allocate a fixed number of tokens for all input clips, ignoring the heterogeneous representation complexity across videos. Such uniform allocation may waste transmission resources on simple clips while providing insufficient representation capacity for complex ones. To address this issue, we develop a reconstruction-proxy-guided adaptive routing module, which determines both the clip-level token budget and the content-priority token positions according to the input video.

	\subsubsection{Motivation for Adaptive Routing}
	
	From the perspective of variable-length source coding \cite{shannon1959coding}, the representation length assigned to a sample should adapt to its information complexity. For a video clip $\bm{x}$ drawn from an underlying distribution $p(\bm{x})$, this principle motivates the ideal relationship
	\begin{equation}
		L
		\propto
		-\log p(\bm{x}),
		\label{eq:ideal_information_budget}
	\end{equation}
	where $L$ denotes the token budget assigned to $\bm{x}$. Therefore, videos with higher representation complexity should be allocated more tokens. However, the true data distribution $p(\bm{x})$ is unknown and difficult to estimate directly.
	
	To obtain a tractable complexity measure, we consider the evidence lower bound (ELBO):
	\begin{equation}
		\mathrm{ELBO}(\bm{x})
		=
		\mathbb{E}_{q(\bm{z}|\bm{x})}
		[\log p(\bm{x}|\bm{z})]
		-
		D_{\mathrm{KL}}
		(q(\bm{z}|\bm{x})\|p(\bm{z})),
		\label{eq:router_elbo}
	\end{equation}
	where $q(\bm{z}|\bm{x})$ and $p(\bm{x}|\bm{z})$ denote the encoding and decoding distributions, respectively. Since $\mathrm{ELBO}(\bm{x})\leq \log p(\bm{x})$, the negative ELBO provides a variational surrogate of the representation complexity:
	\begin{equation}
		c_{\rm ELBO}(\bm{x})
		=
		-\mathrm{ELBO}(\bm{x}).
	\end{equation}
	Based on this complexity measure, we formulate the content-adaptive routing policy as
	\begin{equation}
		L=\mathcal R(\bm{x})
		=
		\delta\left(
		\beta \cdot N \cdot
		\frac{
			c_{\rm ELBO}(\bm{x})
		}{
			\mathbb{E}[c_{\rm ELBO}(\bm{x})]
		}
		\right),
		\label{eq:ideal_elbo_routing}
	\end{equation}
	where $\beta$ denotes the nominal token ratio and $\delta(\cdot)$ maps the continuous budget estimate into an admissible integer token budget. The normalization term provides a relative complexity measure, such that $\beta N$ represents the nominal budget.
	
	\begin{figure}[t]
		\centering
		\includegraphics[width=0.42\textwidth]{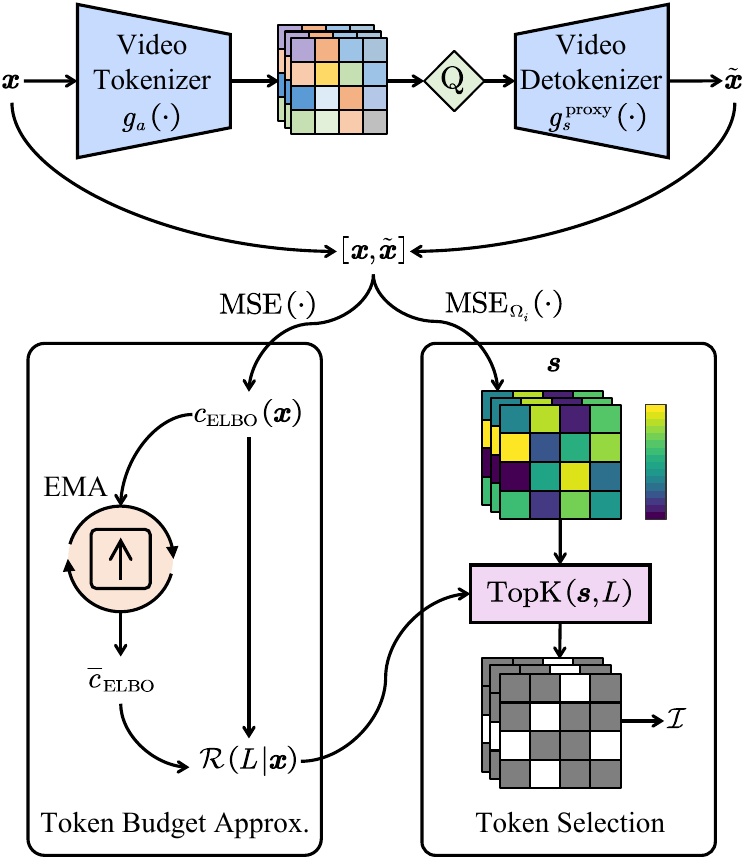}
		\caption{Overview of the reconstruction-proxy-guided adaptive token routing. 
			Given an input video clip $\bm{x}$, the frozen proxy branch, including an independent frozen detokenizer copy, generates the proxy reconstruction $\tilde{\bm{x}}$. The global reconstruction distortion between $\bm{x}$ and $\tilde{\bm{x}}$ is normalized by an EMA complexity reference to determine the adaptive token budget $L$, while the token-aligned reconstruction error map provides localized complexity scores for selecting the token indices $\mathcal{I}$.}
		\label{fig:adaptive_token_routing}
	\end{figure}
	
	\subsubsection{Reconstruction-Proxy-Based Token Budgeting}
	
	As illustrated in Fig.~\ref{fig:adaptive_token_routing}, the same reconstruction proxy is used at two levels: the global reconstruction distortion determines the clip-level token budget, while the localized distortion guides token selection. Specifically, we create an independent copy $g_s^{\rm proxy}$ of the pretrained detokenizer for the proxy branch. Both $g_a$ and $g_s^{\rm proxy}$ remain frozen throughout training. The proxy reconstruction is computed as
	\begin{equation}
		\bm{x}
		\xrightarrow{g_a(\cdot)}
		\bm{y}
		\xrightarrow{Q(\cdot)}
		\bar{\bm{y}}
		\xrightarrow{g_s^{\rm proxy}(\cdot)}
		\tilde{\bm{x}}.
		\label{eq:slide_reconstruction}
	\end{equation}
	
	Inspired by the reconstruction-based approximation in \cite{ye2026infotok}, we use the reconstruction distortion as a tractable estimation of the ELBO-induced complexity:
	\begin{equation}
		c_{\rm ELBO}(\bm{x})
		\approx
		\mathcal{D}(\bm{x},\tilde{\bm{x}})
		=
		\operatorname{MSE}(\bm{x},\tilde{\bm{x}}).
		\label{eq:reconstruction_complexity_proxy}
	\end{equation}
	A larger $c_{\rm ELBO}(\bm{x})$ indicates that the clip is more difficult to represent using the frozen reconstruction proxy and should therefore receive a larger token budget.
	
	To approximate the dataset-level expectation in \eqref{eq:ideal_elbo_routing}, we maintain an exponential moving average (EMA) $\bar c_{\rm ELBO}$ of the estimated complexity during training, which is fixed during inference. The normalized complexity and its shrinkage-adjusted version are computed as
	\begin{equation}
		\rho
		=
		\frac{
			c_{\rm ELBO}(\bm{x})
		}
		{
			\bar c_{\rm ELBO}
		},
		\qquad
		\widetilde{\rho}
		=
		1+\kappa(\rho-1),
		\label{eq:normalized_proxy_complexity}
	\end{equation}
	where $\kappa\in[0,1]$ controls the adaptation strength.
	
	The final token budget is obtained as
	\begin{equation}
		L
		=
		\operatorname{clip}_{[L_{\min},N]}
		\left(
		\operatorname{round}
		(
		\beta N\widetilde{\rho}
		)
		\right),
		\label{eq:practical_routing}
	\end{equation}
	where $L_{\min}$ denotes the minimum token budget to avoid excessive token pruning. Here, $\beta N$ specifies the nominal token budget, while $\widetilde{\rho}$ adjusts the budget according to the representation complexity of each clip.

	\subsubsection{Token-Level Reconstruction-Guided Selection}
	
	In addition to determining the clip-level token budget, the reconstruction proxy provides localized information for selecting informative token positions. Specifically, the reconstruction error map between $\bm{x}$ and $\tilde{\bm{x}}$ is pooled to the spatiotemporal resolution of the latent-token grid. The score of the $i$-th token is defined as
	\begin{equation}
		s_i
		=
		\mathcal{D}_{\Omega_i}
		(\bm{x},\tilde{\bm{x}})
		=
		\operatorname{MSE}_{\Omega_i}
		(\bm{x},\tilde{\bm{x}}),
		\quad
		i=1,\ldots,N,
		\label{eq:token_reconstruction_score}
	\end{equation}
	where $\Omega_i$ denotes the input spatiotemporal region aligned with the $i$-th latent token. The token-level scores are flattened according to the tokenizer ordering to form $\bm{s}=[s_1,\ldots,s_N]$, where larger values indicate higher reconstruction difficulty and content priority.
	Given the routed budget $L$, the selected token index set is obtained by
	\begin{equation}
		\mathcal I
		=
		\operatorname{TopK}(\bm{s},L),
		\label{eq:reconstruction_guided_selection}
	\end{equation}
	where $\operatorname{TopK}(\bm{s},L)$ returns the indices of the $L$ largest token-level complexity scores. The resulting pair $(L,\mathcal I)$ defines the source-side routing decision for each video clip.

	\begin{figure*}[t]
		\centering
		\includegraphics[width=1\textwidth]{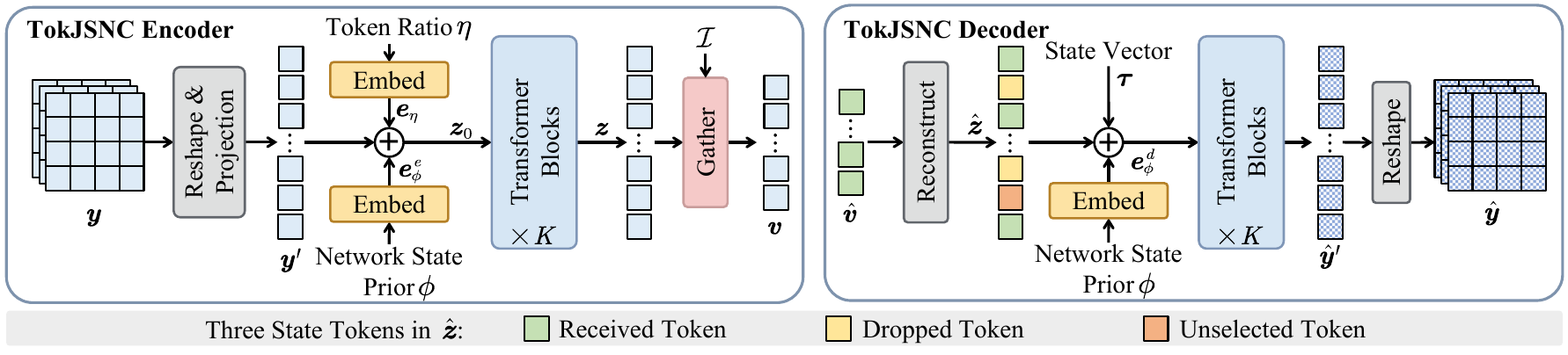}
		\caption{Overview of the proposed network-state-conditioned token encoder and state-aware token decoder. The TokJSNC encoder incorporates the token ratio $\eta$ and shared network state prior $\boldsymbol{\phi}$ to generate network-state-conditioned transport tokens, while the TokJSNC decoder exploits the token state vector $\boldsymbol{\tau}$ and the same network state prior to recover incomplete latent tokens.}
		\label{fig:compressor_decompressor}
	\end{figure*}
	
	\subsection{Network-State-Conditioned Token Encoder and Decoder}
	\label{subsec:network_conditioned_coding}
	Different packet-loss patterns lead to different token availability conditions after transmission. Therefore, beyond token packetization, the transport representation and latent-token recovery process should adapt to the statistical characteristics of the packet-loss network. To this end, we introduce the network-state-conditioned token encoder and state-aware token decoder illustrated in Fig.~\ref{fig:compressor_decompressor}. The shared network state prior $\boldsymbol{\phi}$ provides slow-timescale network information before transmission: the encoder exploits $\boldsymbol{\phi}$ to generate transport representations adapted to different loss conditions, while the decoder combines $\boldsymbol{\phi}$ with the observed token states to recover incomplete latent representations.
	
	\subsubsection{Network State Prior Representation}
	
	Unlike the instantaneous token-reception vector $\mathbf F$, which is only available after transmission, the network state prior $\boldsymbol{\phi}$ characterizes the statistical packet-loss condition shared by the transmitter and receiver before encoding. We consider three representative packet-loss network models:
	\begin{equation}
		\mathcal{M}
		\in
		\{\mathcal{M}_{r},\mathcal{M}_{b},\mathcal{M}_{s}\},
	\end{equation}
	where $\mathcal{M}_{r}$, $\mathcal{M}_{b}$, and $\mathcal{M}_{s}$ represent random, burst, and structured packet-loss processes, respectively. The random model describes memoryless packet loss, the burst model captures temporally correlated losses, and the structured model represents trace-driven or periodic loss patterns.
	
	In addition to the loss model class, the network condition is described by a low-dimensional statistical descriptor $\boldsymbol{\ell}$, which summarizes the average loss rate, burst-length characteristics, and temporal regularity of loss events. A network state encoder generates the shared network state prior:
	\begin{equation}
		\boldsymbol{\phi}
		=
		G_{\theta_{\rm c}}
		(\mathcal{M},\boldsymbol{\ell}),
	\end{equation}
	where $G_{\theta_{\rm c}}(\cdot)$ maps the network profile into a continuous network-state representation. $\boldsymbol{\phi}$ is updated at a slower timescale than individual packet loss and is available at both the transmitter and receiver before transmission. Therefore, it provides statistical network information without revealing the future packet-loss realization.
	
	\subsubsection{TokJSNC Encoder}
	
	Given the latent tokens $\bm y$, the encoder $f_e$ transforms them into transport-oriented representations conditioned on the source-side token ratio $\eta$ and network state prior $\boldsymbol{\phi}$. Specifically, $\bm y$ is first reshaped and projected into the transmission feature space, yielding $\bm y'$. Meanwhile, the token ratio $\eta$ and network state prior $\boldsymbol{\phi}$ are embedded into the same feature dimension, producing $\bm e_{\eta}$ and $\bm e_{\phi}^e$, respectively. The conditioned input representation is obtained as
	\begin{equation}
		\bm z_0
		=
		\bm y'
		+
		\bm e_{\eta}
		+
		\bm e_{\phi}^e.
	\end{equation}
	$\bm z_0$ is then processed by the encoder:
	\begin{equation}
		\bm z
		=
		f_{e}
		(\bm z_0),
	\end{equation}
	where $f_e(\cdot)$ consists of $K$ stacked Transformer \cite{Dosovitskiy2020AnII} blocks and captures token dependencies under the given source and network conditions.
	
	According to the source-side routing result, the transport tokens are extracted:
	\begin{equation}
		\bm v
		=
		\operatorname{Gather}
		(\bm z,\mathcal I).
	\end{equation}
	The transport tokens $\bm v$ are subsequently quantized into discrete symbols for packetized transmission.
	
	\subsubsection{TokJSNC Decoder}
	
	At the receiver, the selected token indices $\mathcal I$ and token-reception vector $\mathbf F$ are jointly used to determine the observation state of each latent-token position. Specifically, each latent position is classified into three mutually exclusive states: received, dropped, and unselected. The state vector $\boldsymbol{\tau}=[\tau_1,\ldots,\tau_N]$ is defined as
	\begin{equation}
		\tau_j
		=
		\begin{cases}
			0, & \exists l:j=\mathcal I_l,F_l=1,\\
			1, & \exists l:j=\mathcal I_l,F_l=0,\\
			-1, & j\notin\{\mathcal I_1,\ldots,\mathcal I_L\}.
		\end{cases}
	\end{equation}
	where $\tau_j=0$, $\tau_j=1$, and $\tau_j=-1$ correspond to received, dropped, and unselected tokens, respectively. These states correspond to different sources of missing information: dropped tokens are lost during transmission, while unselected tokens are removed during source-side routing. Therefore, they should be distinguished during latent-token recovery rather than treated as a unified missing state.
	
	Based on the token states, the receiver constructs the state-aware input representation $\hat{\bm z}$. The received tokens $\hat{\bm v}$ are rearranged to their original latent positions, while dropped and unselected positions are padded with the learnable embeddings $\bm e_{\rm d}$ and $\bm e_{\rm u}$, respectively. Specifically, the $j$-th position of $\hat{\bm z}$ is initialized as
	\begin{equation}
		\hat {\bm z}_j
		=
		\begin{cases}
			\hat{{\bm v}_l},
			&j=\mathcal I_l,F_l=1,\\
			\bm e_{\rm d},
			&j=\mathcal I_l,F_l=0,\\
			\bm e_{\rm u},
			&j\notin\{\mathcal I_1,\ldots,\mathcal I_L\}.
		\end{cases}
	\end{equation}
	
	The token-state vector $\boldsymbol{\tau}$ is directly added as a positional condition. Meanwhile, the network state prior $\boldsymbol{\phi}$ is embedded into the same $d$-dimensional latent feature space, denoted as $\bm e_{\phi}^{d}$. The decoder input is obtained as
	\begin{equation}
		\hat{\bm z}'
		=
		\hat{\bm z}
		+
		\boldsymbol{\tau}
		+
		\bm e_{\phi}^{d}.
	\end{equation}
	Here, $\boldsymbol{\tau}$ is broadcast along the feature dimension to match the $N\times d$ shape of $\hat{\bm z}$.
	
	The complete latent representation is reconstructed by the decoder:
	\begin{equation}
		\hat{\bm y}'
		=
		f_d
		(\hat{\bm z}'),
	\end{equation}
	where $f_d(\cdot)$ consists of $K$ stacked Transformer blocks. The output $\hat{\bm y}'$ is subsequently reshaped into the original spatiotemporal latent-token grid $\hat{\bm y}$.

		\subsection{Spatiotemporal Token Packetization}
	\label{subsec:token_partition}

	After token selection and quantization, the selected transport tokens need to be mapped into packets for transmission. Since packet losses occur at the packet level, all tokens assigned to the same packet become unavailable simultaneously when that packet is lost. Therefore, the packetization strategy determines the correlation structure of token losses and directly affects the difficulty of subsequent latent-token recovery. An effective packetization should distribute tokens across the spatiotemporal domain to avoid concentrated missing regions after packet loss.
	
	Let the latent tokens be organized as a spatiotemporal grid $\mathcal G\in\mathbb{Z}^{t\times h\times w}$, where $t$, $h$, and $w$ denote the temporal and spatial dimensions of the latent-token grid. A packetization strategy divides $\mathcal G$ into multiple spatiotemporal regions:
	\begin{equation}
		\mathcal S
		=
		\{S_1,S_2,\ldots,S_M\},
	\end{equation}
	where each region contains a subset of token positions. The selected transport tokens whose original positions belong to the same region are grouped into one packet:
	\begin{equation}
		P_m
		=
		\{
		\widetilde{v}_i
		|
		(t_i,h_i,w_i)\in S_m
		\},
	\end{equation}
	where $P_m$ denotes the $m$-th transmitted packet. Different packetization strategies therefore lead to different packet-induced missing-token patterns under the same packet-loss process.

\begin{figure}[t]
		\centering
		\includegraphics[width=0.45\textwidth]{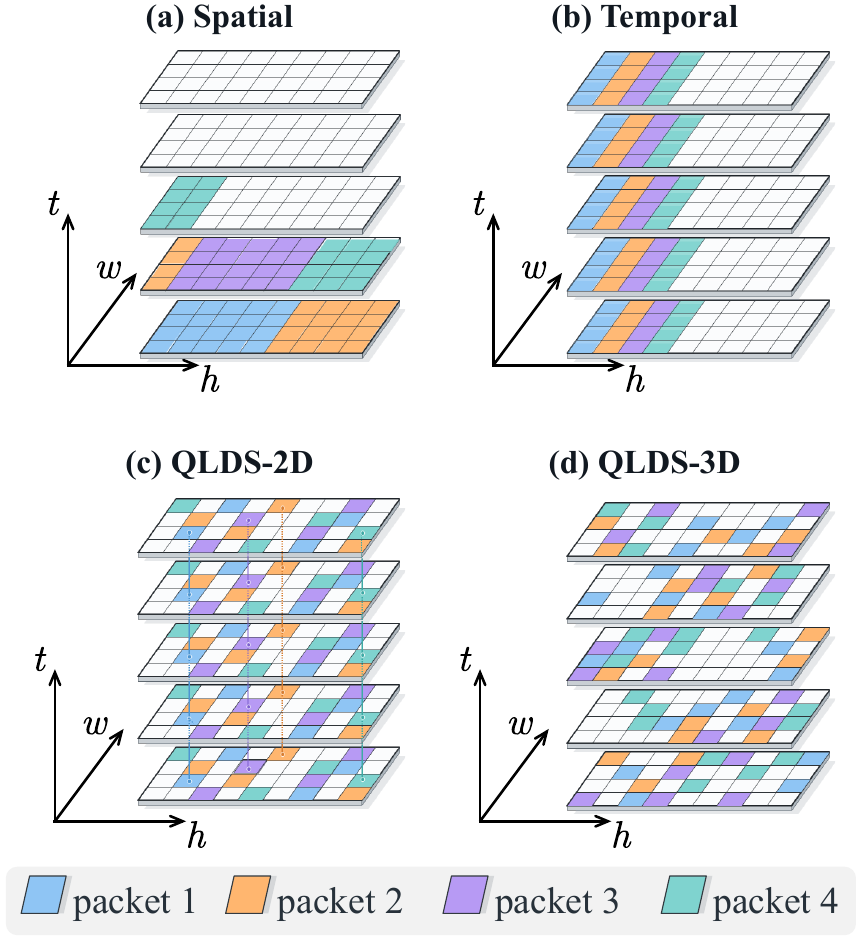}
		\vspace{-0.3cm}
		\caption{Illustration of different spatiotemporal token packetization strategies. Different packetization schemes determine the token grouping within each packet and consequently affect the missing-token patterns under packet losses.
		}
		\vspace{-0.2cm}
		\label{fig:partition}
	\end{figure}

	As illustrated in Fig.~\ref{fig:partition}, we consider several representative packetization strategies. Spatial packetization groups tokens according to spatial locations and preserves local spatial structures, but packet losses may remove concentrated spatial regions. Temporal packetization groups tokens along the temporal dimension and may cause the loss of complete temporal trajectories.
	
	To further reduce token clustering within packets, we adopt the quantized low-discrepancy sequence (QLDS) schedule~\cite{mentzer2023m2t}. QLDS generates a deterministic traversal order with uniform coverage over the token space. The original QLDS is designed for two-dimensional token grids; in this work, we extend it to spatiotemporal latent-token grids by introducing temporal dispersion, resulting in two variants. QLDS-2D performs low-discrepancy sampling within spatial dimensions, while QLDS-3D further incorporates the temporal dimension to distribute tokens across the complete $(t,h,w)$ space.
	
	These packetization strategies generate different token availability patterns after packet losses, which directly influence the state-aware token recovery process at the receiver.

	\subsection{Multi-Stage Training Strategy}
	\label{subsec:training_strategy}
	
	We initialize the video tokenizer from a pretrained Open-MAGVIT2 model \cite{luo2024open}. Following Open-MAGVIT2, our model employs lookup-free quantization (LFQ)\cite{yu2024language} to obtain discrete latent tokens. LFQ uses a parameter-free binary mapping and therefore contains no trainable codebook. The pretrained tokenizer encoder $g_a$ is kept frozen during the following three-stage training. The routing proxy uses an independent copy $g_s^{\rm proxy}$ of the pretrained detokenizer, which also remains frozen and is not shared with the reconstruction detokenizer $g_s$ optimized during training.
	
	\textbf{Stage I: Packet-loss-free initialization.}
	We first train without packet loss and optimize the TokJSNC encoder $f_e$, TokJSNC decoder $f_d$, and Detokenizer $g_s$. In addition to pixel and perceptual reconstruction losses, we retain the LFQ entropy and commitment objectives:
	\begin{equation}
		\mathcal L_{\rm I}
		=
		\mathcal L_{1}
		+ \lambda_{\rm p}\mathcal L_{\rm LPIPS}
		+ \lambda_{\rm e}\mathcal L_{\rm LFQ\text{-}ent}
		+ \lambda_{\rm c}\mathcal L_{\rm LFQ\text{-}com}.
		\label{eq:stage1_training_loss}
	\end{equation}
	Here, $\mathcal L_{1}$ denotes the mean absolute error (MAE), $\mathcal L_{\rm LPIPS}$ denotes the LPIPS loss \cite{lpips}, and $\mathcal L_{\rm LFQ\text{-}ent}$ and $\mathcal L_{\rm LFQ\text{-}com}$ denote the LFQ entropy and commitment losses, respectively.
	We use $\lambda_{\rm p}=0.2$, $\lambda_{\rm e}=0.1$, and $\lambda_{\rm c}=0.25$. The LFQ auxiliary losses are computed from the quantizer input and regularize the output of the TokJSNC encoder before binary quantization. We first use a full-token warm-up and then gradually reduce $\beta$ from $1.0$ to the target value, such as $0.8$.
	
	\textbf{Stage II: packet-loss-aware training.}
	Starting from the Stage I checkpoint, we enable simulated random, burst, and structured packet-loss processes. In addition to $f_e$, $f_d$, and $g_s$, we introduce and jointly train the network state encoder $G_{\theta_{\rm c}}$ in this stage. The reconstruction objective remains $\mathcal L_{\rm I}$, while $\beta$ is fixed. The packet-loss model and its statistical descriptors are sampled during training, allowing $G_{\theta_{\rm c}}$ to learn the network-state prior and enabling $f_e$ and $f_d$ to learn network-state-conditioned transmission and latent-token recovery.

	\textbf{Stage III: adversarial receiver refinement.}
	In this stage, we train the TokJSNC decoder $f_d$, the Detokenizer $g_s$, and discriminator (a three-layer spatiotemporal PatchGAN \cite{li2016precomputed} with 64 base channels), while all other modules are kept fixed.
	The generator and discriminator objectives are
	\begin{equation}
		\begin{aligned}
			\mathcal L_{\rm G}
			&=
			\mathcal L_{1}
			+ \lambda_{\rm p}\mathcal L_{\rm LPIPS}
			+ \lambda_{\rm adv}\mathcal L_{\rm adv},
			\\
			\mathcal L_{\rm D}
			&=
			\lambda_{\rm d}
			\left(
			\mathcal L_{\rm hinge}
			+ \lambda_{\rm LeCAM}\mathcal L_{\rm LeCAM}
			\right).
		\end{aligned}
		\label{eq:stage3_training_loss}
	\end{equation}
	Here, $\mathcal L_{\rm adv}$ denotes the generator adversarial loss, $\mathcal L_{\rm hinge}$ denotes the hinge discriminator loss \cite{miyato2018spectral}, and $\mathcal L_{\rm LeCAM}$ denotes the LeCAM regularization loss \cite{tseng2021regularizing}.
	We use $\lambda_{\rm p}=1.0$, $\lambda_{\rm adv}=0.3$, $\lambda_{\rm d}=0.8$, and $\lambda_{\rm LeCAM}=0.05$.
	
	\begin{figure}[t]
		\centering
		\includegraphics[width=0.49\textwidth]{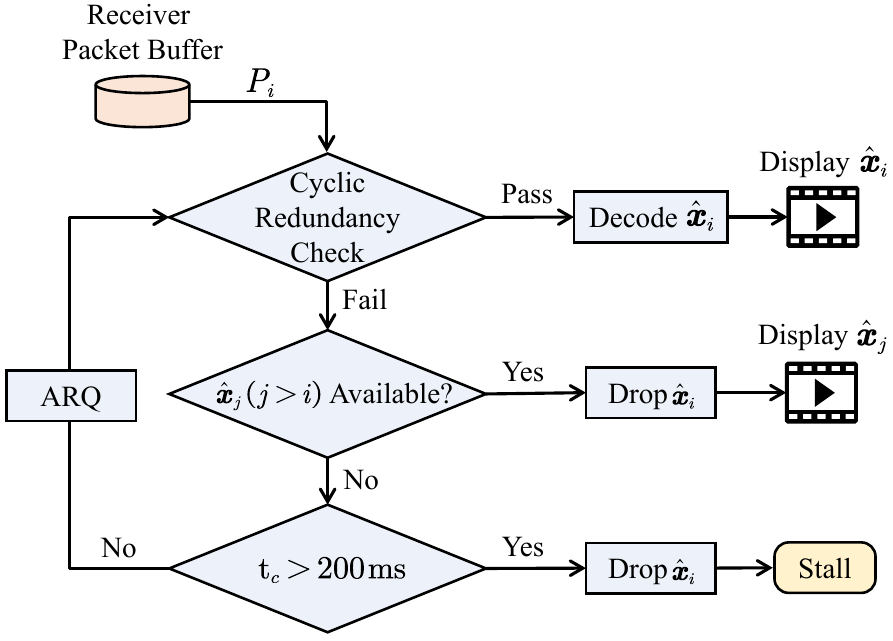}
		\vspace{-0.5cm}
		\caption{Receiver-side stall detection process for WebRTC-based methods. A stall is determined by considering packet recovery, decoded future-frame availability, and playback deadlines, where frame dropping can avoid stalls when subsequent frames are available.}
		\label{fig:stall_logic}
	\end{figure}
	
	\section{Experiments} \label{experiments}
	
	\subsection{Experimental Setup}
	
	\subsubsection*{\normalfont\bfseries Implementation Details}
	We initialize the video tokenizer from a pretrained Open-MAGVIT2 checkpoint configured with a temporal compression factor of 4 and spatial compression factors of $8\times8$. The LFQ code uses 18 binary dimensions, so each quantized token is represented by an 18-bit symbol before packetization. The TokJSNC encoder and decoder each contain 8 Transformer blocks. The hidden dimension is 256, the number of attention heads is 32, and the MLP hidden dimension is 512. The network state encoder maps the loss-process identifier and 3 statistical descriptors into a 64-dimensional network-state representation through a two-layer MLP. For adaptive routing, the reconstruction-proxy EMA decay is set to 0.99, the minimum token ratio is $1/16$, and the token-allocation shrink factor is 0.3. We train the models with target routing ratios $\beta\in\{0.5,0.65,0.8\}$.
	
	\subsubsection*{\normalfont\bfseries Training Details}
	We train TokJSNC on the 32-frame version of the original Vimeo dataset~\cite{li2024neural} using four NVIDIA RTX 4090 GPUs. Each training sample contains 17 consecutive frames, which are randomly cropped to $128\times128$. Random horizontal and vertical flipping is applied for data augmentation. The per-GPU batch size is 12, resulting in a total batch size of 48. The learning rates of the generator and discriminator are both set to $1\times10^{-5}$. We use mixed-precision training, clip the gradient norm to 1.0, and set the random seed to 42. During packet-loss-free training, the router first uses all tokens for 5,000 steps, after which $\beta$ is linearly decreased from 1.0 to target value over 10,000 steps. During packet-loss-aware training, $\beta$ is fixed at 0.8 and the packet-loss ratio is uniformly sampled from $[0,0.3]$.
	
	\subsubsection*{\normalfont\bfseries Evaluation Datasets}
	
	We use the UVG dataset~\cite{mercat2020uvg} for both offline codec evaluation and end-to-end streaming evaluation. The end-to-end streaming results are reported on the seven UVG sequences. Each test video is resized to $256\times480$ and evaluated at 30 frames per second. For offline codec evaluation, we use the first 153 frames of each video, corresponding to nine consecutive GOPs of 17 frames. For end-to-end streaming evaluation, each run contains a 300-frame warm-up period followed by 153 measured frames, allowing the transport controller and pacing process to reach a stable operating state before collecting statistics.
	
	\subsubsection*{\normalfont\bfseries Compared Methods}
	We organize the evaluated methods into two groups. The WebRTC-based \cite{w3cwebrtc10} baselines include the conventional video codecs H.264/AVC~\cite{wiegand2003overview} and H.265/HEVC~\cite{sze2014high}, the token-based Cosmos tokenizer~\cite{nvidia2025cosmos}, and the real-time neural video codec DCVC-RT~\cite{jia2025towards}, all of which are integrated with WebRTC for packet transport and loss recovery. The loss-resilient methods include GRACE~\cite{cheng2024grace} and TokJSNC, both of which can reconstruct each frame from incomplete received representations and are evaluated using their native pipelines rather than the WebRTC loss-recovery path. H.264 and H.265 rate ladders are encoded without B-frames, using one reference frame, a 17-frame GOP, and MTU-aware packetization. DCVC-RT uses a seven-point rate ladder from 140 to 260~kbps, Cosmos uses two representations at 163 and 271~kbps, and GRACE is evaluated using its lowest-bitrate released model. For the token-based representations, Cosmos and TokJSNC directly convert quantized token symbols into bitstreams for packetization without entropy coding. TokJSNC operates at an application-layer rate of approximately 270~kbps.
	
	\subsubsection*{\normalfont\bfseries Evaluation Metrics}
	We report MS-SSIM~\cite{wang2003multiscale}, where higher values indicate better reconstruction quality, together with LPIPS~\cite{lpips}, DISTS~\cite{dists}, and FID~\cite{fid}, where lower values indicate better perceptual quality. MS-SSIM, LPIPS, and DISTS are computed for individual frames and averaged over all evaluated frames. FID measures distribution-level reconstruction fidelity and is computed using overlapping $256\times256$ patches extracted from the reconstructed and reference videos.

	For streaming-continuity evaluation, we apply the same stall-detection criterion to all methods. 
	A frame is considered successfully recovered only if it is available within the 200~ms playback deadline and its RGB PSNR is no lower than 18~dB. 
	Otherwise, the frame is treated as a decoding failure. 
	A continuous decoding-failure interval lasting at least 200~ms is counted as one stall event.
	As illustrated in Fig.~\ref{fig:stall_logic}, for methods integrated with WebRTC, each received packet $P_i$ is first verified using a cyclic redundancy check. 
	If the check fails, retransmission is requested through automatic repeat request (ARQ) protocol, and the receiver waits for the retransmitted packet until the playback deadline. 
	For loss-resilient video streaming methods such as GRACE and TokJSNC, which support reconstruction from incomplete packet reception, retransmission is disabled and frames are reconstructed directly from the packets available before the playback deadline.
	Despite these differences in recovery mechanisms, all methods are evaluated using the same 200~ms deadline and 18~dB RGB-PSNR threshold.

	For quantitative evaluation, we report the stall ratio, average stall duration, number of stall events, and average playback frame rate. The average stall duration is defined as
	\begin{equation}
		T_{\rm stall}^{\rm avg}
		=
		\frac{1}{N_{\rm stall}}
		\sum_{i=1}^{N_{\rm stall}} T_{\rm stall}^{(i)},
	\end{equation}
	where $N_{\rm stall}$ denotes the number of stall events and $T_{\rm stall}^{(i)}$ represents the duration of the $i$-th stall event. The stall ratio is calculated as
	\begin{equation}
		R_{\rm stall}
		=
		\frac{
			\sum_{i=1}^{N_{\rm stall}}T_{\rm stall}^{(i)}
		}
		{T_{\rm total}},
	\end{equation}
	where $T_{\rm total}$ denotes the total playback duration. The number of stall events is directly given by $N_{\rm stall}$. To complement the stall-based metrics, we further introduce the average playback frame rate, which measures the number of frames displayed per second (FPS) over the nominal video duration.
	
	\begin{figure*}[t]
		\centering
		\includegraphics[width=\textwidth]{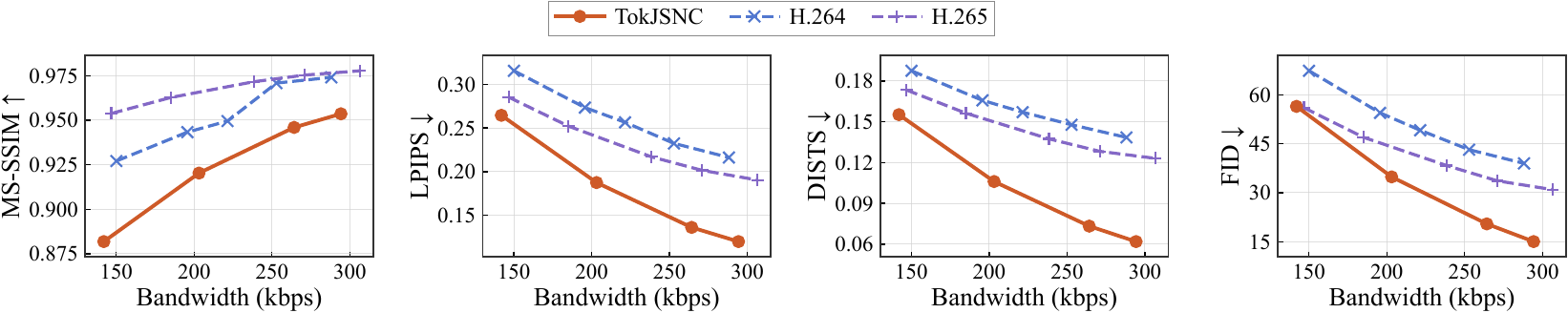}
		\vspace{-0.5cm}
		\caption{Codec reconstruction performance under loss-free delivery, without packetization or network transport. The four TokJSNC operating points, ordered by increasing rate, use $\beta\in\{0.5,0.65,0.8,1.0\}$, while H.264 and H.265 use multiple QP settings.}
		\label{fig:clean_rd_results}
	\end{figure*}
	
	\begin{figure*}[t]
		\centering
		\includegraphics[width=\textwidth]{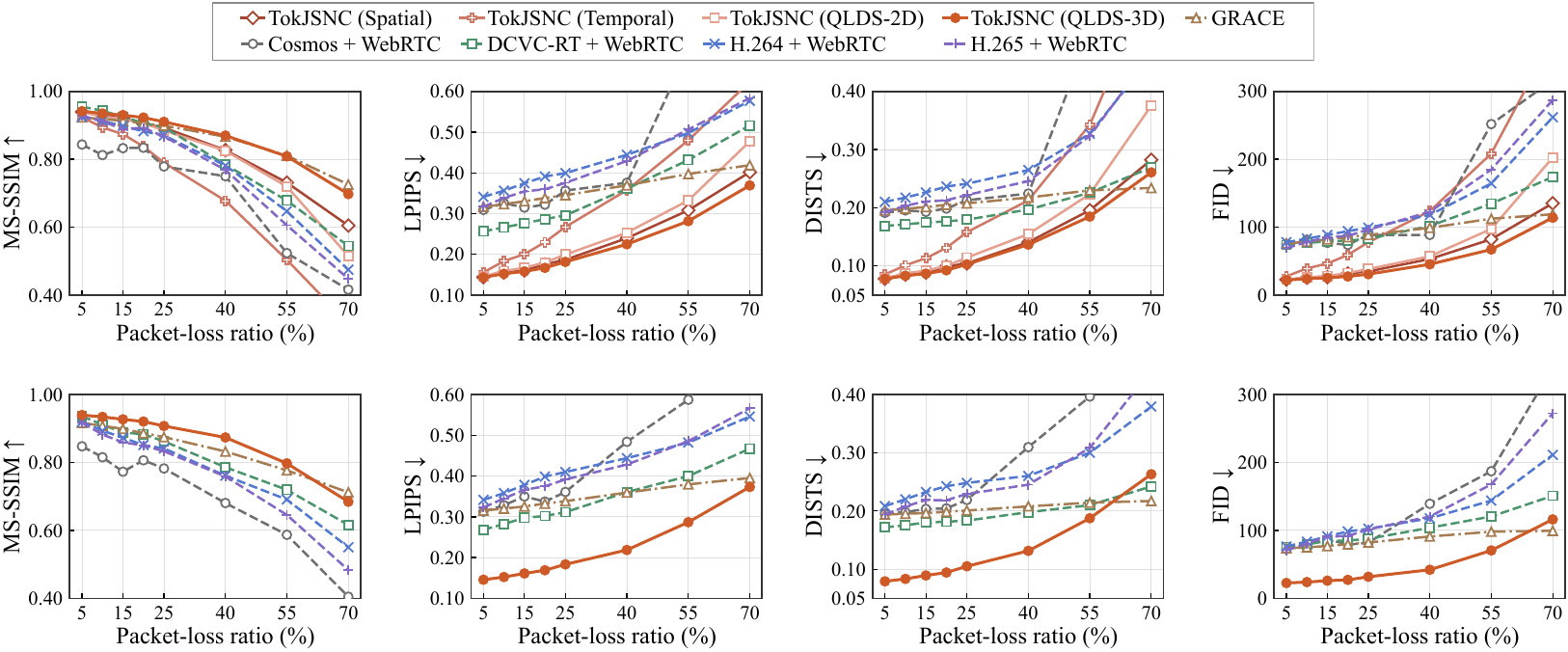}
		\vspace{-0.5cm}
		\caption{End-to-end reconstruction quality under random and burst packet losses from 5\% to 70\%. The upper and lower rows correspond to random and burst packet loss, respectively. From left to right, the columns report MS-SSIM, LPIPS, DISTS, and FID. The TokJSNC variants differ in their token-to-packet packetization strategies. Results are averaged over ten random seeds.}
		\vspace{-0.2cm}
		\label{fig:shared_quality_results}
	\end{figure*}
	
	\subsubsection*{\normalfont\bfseries Network Setup}
	
	We conduct the end-to-end evaluation using the SparkRTC/ns-3 packet-level simulator~\cite{meng2022achieving}, which models frame generation, packetization, pacing, congestion control, loss recovery, playback deadlines, and decoding. Unless otherwise stated, the bottleneck capacity is 1000~kbps, the application-layer rate budget is 275~kbps, the one-way propagation delay is 20~ms, the packet payload is 1200~bytes, and the frame deadline is 200~ms. The source frame rate is fixed at 30~fps. The WebRTC-based baselines use Google Congestion Control (GCC), a pacing factor of 4, adaptive FEC, duplicate-ACK retransmission, and cross-frame FEC with a five-packet group and a maximum waiting time of 100~ms. H.264 and H.265 use their causal WebRTC recovery paths, while Cosmos and DCVC-RT are evaluated without picture-loss indication (PLI). GRACE is evaluated with its native loss-resilient mechanism. TokJSNC does not rely on retransmission, FEC, or PLI, and instead reconstructs incomplete token representations through its error-resilient decoder. To ensure comparable operating rates, TokJSNC uses an approximately 270~kbps application-layer representation and is paced within the same 275~kbps budget used by the baselines; after transport overhead, its measured total wire rate averages approximately 284~kbps. All methods are evaluated under the same bottleneck link, packetization constraints, playback deadlines, and packet-loss traces. TokJSNC is conditioned only on coarse loss statistics available before transmission, such as the average loss rate and burst characteristics, without access to future packet-loss realizations. The reported transmission rate is measured from transmitted packet bytes and includes transport overhead and recovery traffic rather than the nominal codec rate.

	\begin{figure*}[t]
		\centering
		\includegraphics[width=\textwidth]{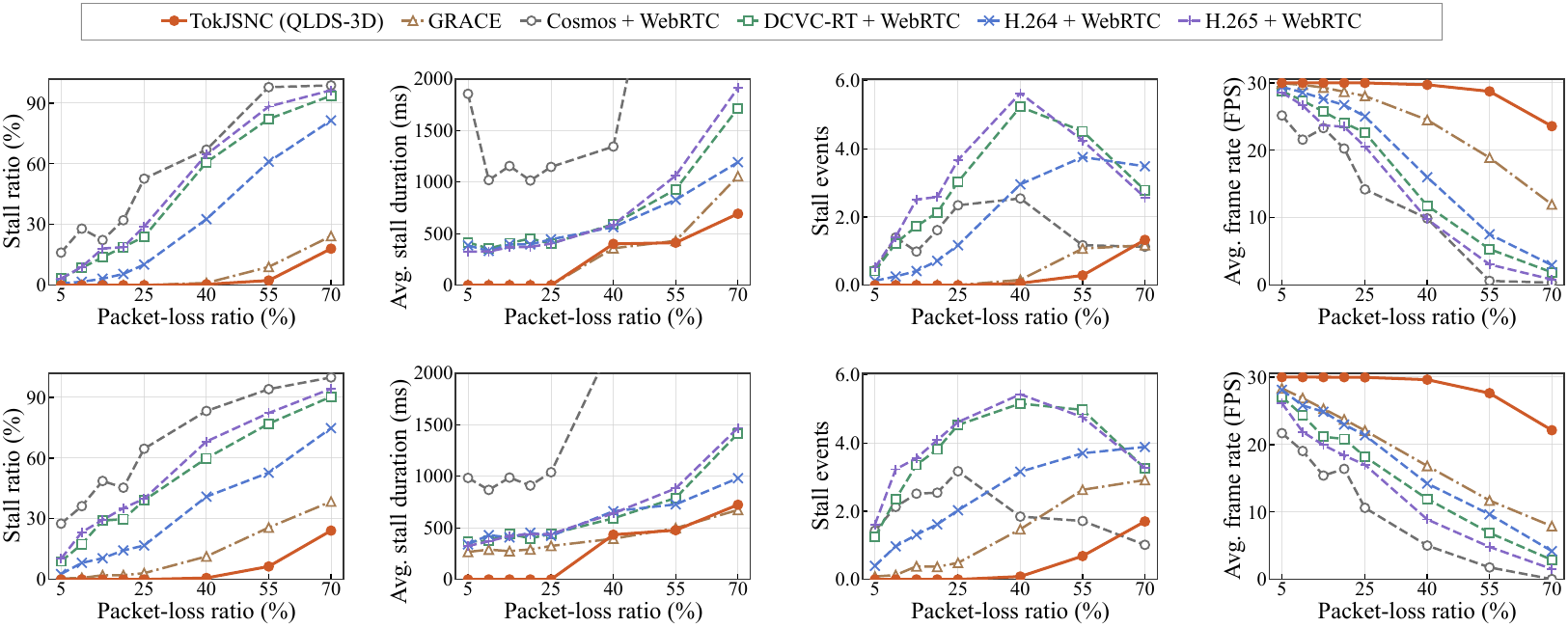}
		\vspace{-0.5cm}
		\caption{Playback-continuity performance under random and burst packet losses. The upper and lower rows correspond to random and burst packet loss, respectively. From left to right, the columns report stall ratio, average stall duration, number of stall events, and average playback frame rate. TokJSNC uses QLDS-3D. Results are averaged over ten random seeds and seven videos, with 153 frames evaluated per video.}
		\label{fig:shared_stall_summary_results}
	\end{figure*}

	\begin{figure}[!t]
		\centering
		\setlength{\abovecaptionskip}{0pt}
		\includegraphics[width=0.85\columnwidth]{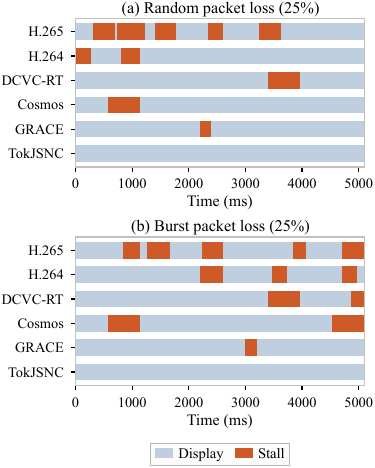}
		\caption{Representative per-frame playback timelines for the Beauty sequence under 25\% random and burst packet loss. H.264, H.265, DCVC-RT, and Cosmos use WebRTC pipelines, whereas GRACE uses its native loss-resilient pipeline. Blue and orange denote displayed and stalled intervals, respectively.}
		\vspace{-0.2cm}
		\label{fig:stall_timeline}
	\end{figure}
	
	\subsection{Codec Performance under Loss-Free Delivery}
	
	We first evaluate the intrinsic reconstruction performance of TokJSNC under loss-free delivery, excluding packetization, network transport, and packet-recovery mechanisms. As shown in Fig.~\ref{fig:clean_rd_results}, increasing the nominal token ratio $\beta$ consistently improves the reconstruction quality across all four metrics, providing controllable operating points at different measured rates. At medium- and high-rate operating points, TokJSNC achieves lower LPIPS, DISTS, and FID than H.264 and H.265 at comparable rates, indicating improved perceptual similarity and distributional realism. In contrast, its MS-SSIM remains lower than those of conventional codecs. This difference reflects the perception-oriented optimization of TokJSNC, where perceptual refinement improves visual plausibility but does not necessarily maximize pixel-level structural similarity. 
	
	\begin{figure*}[t]
		\centering
		\includegraphics[width=\textwidth]{%
			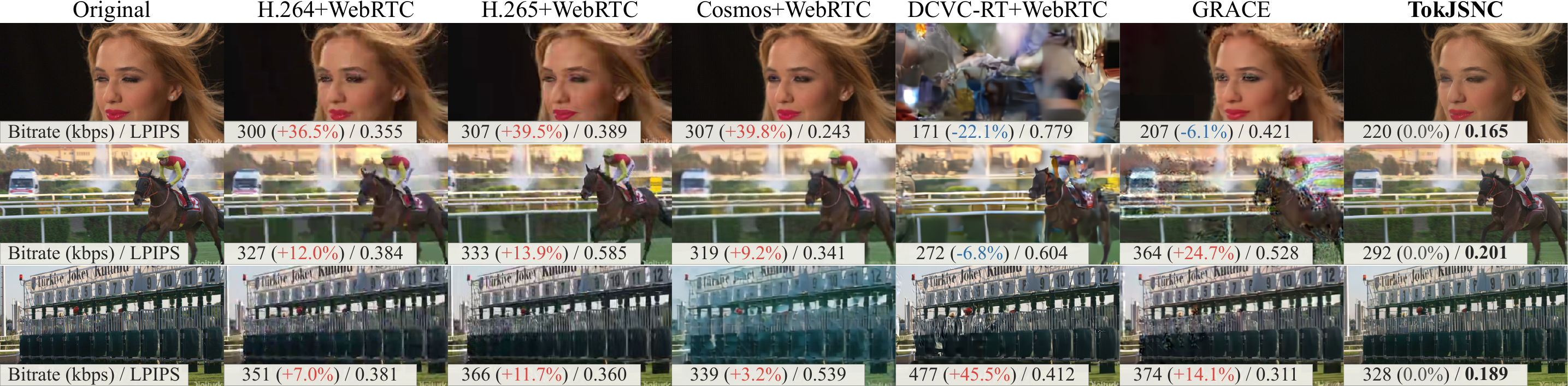%
		}
		\caption{Visual comparison of representative reconstructed frames from three sequences under 25\% random packet loss. Each reconstruction is annotated with its measured end-to-end transmission bitrate, the relative bitrate difference from TokJSNC for the same sequence (TokJSNC: 0.0\%), and LPIPS.}
		\vspace{-0.2cm}
		\label{fig:qualitative_results}
	\end{figure*}
	
	\begin{figure*}[t]
		\centering
		\includegraphics[width=0.98\textwidth]{%
			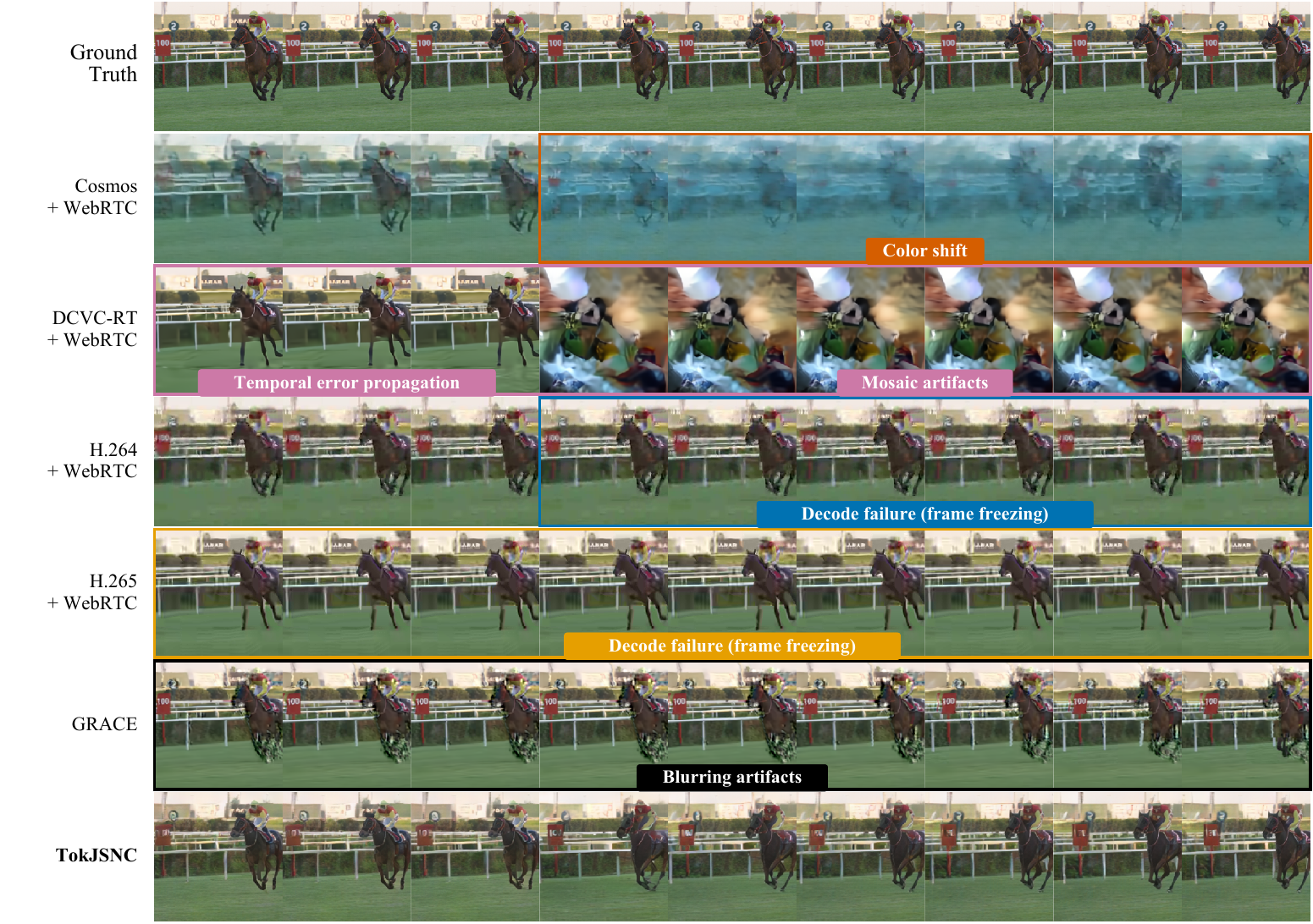%
		}
		\caption{Consecutive-frame reconstruction comparison for the Jockey sequence under 40\% random packet loss. Colored boxes and labels highlight representative visual and temporal failure modes. TokJSNC maintains stable reconstruction throughout the illustrated interval.}
	\vspace{-0.2cm}
	\label{fig:temporal_visualization}
\end{figure*}

\subsection{End-to-End Performance under Packet Loss}

We next evaluate the complete packetized transmission system under the network configuration in Section~IV-A. All methods are evaluated under the same network conditions and playback deadline. Since recovery mechanisms introduce different overheads, the results are reported as end-to-end system comparisons rather than equal-rate codec comparisons. Reconstruction quality and playback continuity are evaluated under both random and burst packet losses.

\subsubsection*{\normalfont\bfseries Reconstruction Quality}

The upper row of Fig.~\ref{fig:shared_quality_results} compares the end-to-end reconstruction quality under random packet loss. As the packet-loss ratio increases, all methods experience quality degradation, but their degradation behaviors differ substantially. The WebRTC-based baselines deteriorate rapidly under moderate-to-severe packet losses, whereas GRACE and TokJSNC exhibit considerably smoother degradation by reconstructing video content from partially received representations. Compared with GRACE, TokJSNC with QLDS-3D achieves comparable MS-SSIM and generally lower LPIPS, DISTS, and FID over most of the evaluated loss range. Overall, these results demonstrate that TokJSNC maintains a favorable balance between structure-oriented fidelity and perceptual and distributional quality under incomplete packet reception.

The comparison among the TokJSNC variants further demonstrates the importance of token-to-packet packetization. Temporal packetization degrades most rapidly because the tokens grouped into one packet are concentrated along the temporal dimension, causing each packet loss to remove a strongly correlated subset of temporal information. Spatial packetization and QLDS-2D alleviate this concentration, while QLDS-3D consistently provides the strongest robustness by distributing the tokens assigned to each packet across all three spatiotemporal dimensions. Consequently, packet loss produces spatially and temporally dispersed missing observations rather than concentrated missing regions. Moreover, since the packet-loss ratio used during training is sampled only from $[0,0.3]$, the smooth degradation observed up to 70\% loss suggests that TokJSNC remains robust well beyond the packet-loss range encountered during training.

The lower row of Fig.~\ref{fig:shared_quality_results} extends the evaluation to time-correlated burst packet loss, with TokJSNC using QLDS-3D. Burst losses introduce temporally correlated packet erasures and therefore create more challenging reconstruction conditions than independent losses. Nevertheless, TokJSNC maintains a smooth quality-degradation trend across the evaluated loss range, achieving MS-SSIM that is competitive with or superior to the baselines, together with generally lower LPIPS, DISTS, and FID. These results demonstrate that the reconstruction robustness of TokJSNC extends beyond independent packet losses to temporally correlated burst conditions.

\begin{table}[t]
	\centering
	\caption{Model complexity and processing latency.}
	\label{tab:complexity_latency}
	\begin{threeparttable}
		\small
		\setlength{\tabcolsep}{1.0pt}
		\renewcommand{\arraystretch}{1.05}
		\begin{tabular*}{\columnwidth}{@{}l@{\extracolsep{\fill}}cccc@{}}
			\toprule
			Method
			& Params (M)
			& Enc. (ms)
			& Dec. (ms)
			& Total (ms) \\
			\midrule
			TokJSNC (Ours)   & 379.3 & 31.3 & 32.4 & 63.7 \\
			Cosmos + WebRTC  & 104.7 &  1.09 &  2.76 &  3.85 \\
			DCVC-RT + WebRTC &  66.3 &  4.71 &  4.33 &  9.04 \\
			GRACE            &   4.75 & 34.2 &  4.33 & 38.5 \\
			H.264 + WebRTC   & N/A    &  2.03 &  0.49 &  2.52 \\
			H.265 + WebRTC   & N/A    &  7.11 &  0.74 &  7.85 \\
			\bottomrule
		\end{tabular*}
		\begin{tablenotes}[flushleft]
			\footnotesize
			\item \textit{Note:} Encoding and decoding latencies are averaged per source frame, and total latency is their sum. The precomputed TokJSNC complexity proxy is excluded from online encoding latency. Parameter counts are not applicable to conventional codecs.
		\end{tablenotes}
	\end{threeparttable}
	\vspace{-0.2cm}
\end{table}

\subsubsection*{\normalfont\bfseries Playback Continuity}

Fig.~\ref{fig:shared_stall_summary_results} compares playback continuity under random and burst packet losses. Across the evaluated packet-loss range, TokJSNC achieves the lowest or tied-lowest stall ratio while maintaining relatively short and infrequent interruptions under both loss models. Its advantage becomes more evident under severe packet loss, where the competing schemes experience rapidly increasing playback disruption. Notably, this continuity advantage is achieved without relying on FEC or retransmission.

At the same average packet-loss ratio, burst loss generally causes more severe playback disruption than random loss, as reflected by a higher stall ratio and a lower average playback frame rate. The average playback frame rate complements the stall metrics because the receiver may discard the current frame and advance to an available future frame, thereby avoiding a stall at the cost of displaying fewer frames. TokJSNC maintains both a low stall ratio and a high average playback frame rate under the two loss models, indicating that its continuity advantage is not obtained through aggressive frame skipping. At 70\% packet loss, TokJSNC sustains average frame rates of 23~fps under random loss and 22~fps under burst loss, whereas the strongest baseline reaches only 12 and 8~fps, respectively. Moreover, the number of stall events is not strictly monotonic with the loss ratio, since consecutive decoding failures may merge into fewer but longer interruptions.

Fig.~\ref{fig:stall_timeline} provides a representative temporal view of the aggregate playback results. Under 25\% random packet loss, the competing methods exhibit isolated or intermittent stall intervals. Under burst packet loss, the interruptions generally become more temporally clustered because consecutive packet losses affect neighboring playback intervals. In contrast, TokJSNC maintains uninterrupted playback throughout the illustrated 5-second interval under both loss models. This example does not imply loss-free reconstruction; rather, it visually confirms that the lower stall ratio of TokJSNC corresponds to fewer missed playback deadlines, consistent with the aggregate results in Fig.~\ref{fig:shared_stall_summary_results}.

\begin{table}[t]
	\centering
	\caption{Rate-matched ablation study of network-state prior under different packet-loss networks.}
	\label{tab:loss_process_channel_ablation}
	\begin{threeparttable}
		\small
		\setlength{\tabcolsep}{1.0pt}
		\renewcommand{\arraystretch}{1.08}
		\begin{tabular*}{\columnwidth}{@{}l@{\extracolsep{\fill}}lcccc@{}}
			\toprule
			Model & Lossy Network & MS-SSIM $\uparrow$ & LPIPS $\downarrow$
			& DISTS $\downarrow$ & FID $\downarrow$ \\
			\midrule
			\multirow{4}{*}{TokJSNC}
			& R-Random
			& \textbf{0.923} & \textbf{0.165} & \textbf{0.090} & \textbf{26.4} \\
			& B-Burst
			& \textbf{0.919} & \textbf{0.168} & \textbf{0.094} & \textbf{27.9} \\
			& S-Structured
			& \textbf{0.912} & \textbf{0.173} & \textbf{0.095} & \textbf{29.3} \\
			& Mahimahi$^{*}$
			& \textbf{0.916} & \textbf{0.170} & \textbf{0.095} & \textbf{27.8} \\
			\midrule
			\multirow{4}{*}{\makecell{w/o\\Network\\Prior}}
			& R-Random
			& 0.904 & 0.182 & 0.096 & 34.9 \\
			& B-Burst
			& 0.891 & 0.189 & 0.103 & 37.8 \\
			& S-Structured
			& 0.899 & 0.188 & 0.101 & 37.7 \\
			& Mahimahi$^{*}$
			& 0.878 & 0.199 & 0.114 & 40.2 \\
			\bottomrule
		\end{tabular*}
		\begin{tablenotes}[flushleft]
			\footnotesize
			\item \textit{Note:} Both models operate at approximately 264~kbps. Results use the first 153 frames of seven UVG videos. $^{*}$Mahimahi denotes a measured trace with approximately 20\% packet loss, replayed at the latent-token level.
		\end{tablenotes}
	\end{threeparttable}
	\vspace{-0.2cm}
\end{table}

\subsubsection*{\normalfont\bfseries Visual Quality Comparison}
Fig.~\ref{fig:qualitative_results} presents representative frame-level reconstructions under 25\% random packet loss, annotated with the measured transmission bitrate and LPIPS. H.264 and H.265 exhibit blocking and structural distortions, DCVC-RT suffers from severe mosaic artifacts, Cosmos introduces color shifts and local-detail degradation, and GRACE produces noticeably blurred results. In contrast, TokJSNC preserves more coherent structures and fine details, achieving the lowest LPIPS in all illustrated examples while operating at a lower transmission rate than most baselines.

Fig.~\ref{fig:temporal_visualization} compares consecutive reconstructed frames and reveals distinct temporal failure modes: persistent color shifts in Cosmos, error propagation and mosaic artifacts in DCVC-RT, decoding-failure-induced frame freezing in H.264 and H.265, and persistent blurring in GRACE. TokJSNC maintains stable color, scene structure, and motion progression without visible error propagation or frame freezing, demonstrating superior perceptual quality and temporal consistency.

\subsection{Model Complexity and Processing Latency}

Table~\ref{tab:complexity_latency} compares model complexity and processing latency. TokJSNC introduces greater model and computational overhead than the conventional and lightweight neural baselines, primarily because of its loss-resilient token recovery. Overall, the results reveal the computational tradeoff associated with the improved reconstruction robustness and playback continuity of TokJSNC.

\subsection{Ablation Studies}

\begin{table}[t]
	\centering
	\caption{Ablation study of adaptive token budgeting.}
	\label{tab:content_aware_token_ablation}
	\begin{threeparttable}
		\small
		\setlength{\tabcolsep}{2.5pt}
		\renewcommand{\arraystretch}{1.05}
		\begin{tabular*}{\columnwidth}{@{}l@{\extracolsep{\fill}}cccc@{}}
			\toprule
			Method
			& MS-SSIM $\uparrow$
			& LPIPS $\downarrow$
			& DISTS $\downarrow$
			& FID $\downarrow$ \\
			\midrule
			TokJSNC
			& \textbf{0.946}
			& \textbf{0.136}
			& \textbf{0.073}
			& \textbf{20.5} \\
			
			w/o Adaptive Budget
			& 0.939
			& 0.142
			& 0.078
			& 27.3 \\
			\bottomrule
		\end{tabular*}
		\begin{tablenotes}[flushleft]
			\footnotesize
			\item \textit{Note:} Both variants retain the same token-selection strategy. The ablated variant replaces the content-adaptive clip-level budget with a fixed token budget while matching the average transmission rate of the full model.
		\end{tablenotes}
	\end{threeparttable}
\end{table}

\subsubsection*{\normalfont\bfseries Network State Prior}

Table~\ref{tab:loss_process_channel_ablation} evaluates the effectiveness of the shared network-state prior by comparing TokJSNC with a variant without network prior conditioning under rate-matched conditions. The complete model consistently improves reconstruction quality across random, burst, structured, and measured loss patterns. The performance gain is particularly evident under the measured Mahimahi trace \cite{netravali2015mahimahi}, where packet-loss characteristics deviate from simple synthetic models. These results demonstrate that incorporating statistical packet-loss characteristics enables the JSNC encoder and decoder to adapt to diverse network conditions and improves robustness under practical loss processes.

\subsubsection*{\normalfont\bfseries Adaptive Token Budgeting}

Table~\ref{tab:content_aware_token_ablation} evaluates the reconstruction-proxy-based adaptive token budgeting mechanism under matched average transmission rates. Both variants retain reconstruction-guided token selection, while the ablated variant assigns a fixed token budget to every clip instead of adapting the budget according to the clip-level reconstruction-derived complexity. Adaptive budgeting consistently improves all four reconstruction metrics, with the largest improvement observed in FID. These results show that reallocating representation capacity across clips with different content complexities improves the utilization of a limited average token budget. 

Together, the two ablations validate the complementary roles of content-adaptive token budgeting and network-state conditioning in TokJSNC. The former improves representation efficiency by adapting the clip-level token budget to content complexity, while the latter improves reconstruction robustness across diverse packet-loss conditions.

\section{Conclusion}  \label{conclusion}
This paper has presented TokJSNC, a new tokenized joint source-network coding framework for loss-resilient video streaming. TokJSNC combines content-adaptive token routing, network-state-conditioned token coding, and spatiotemporal token packetization to reconstruct video from incomplete packet reception without relying on FEC or retransmission. Experiments demonstrate improved perceptual reconstruction quality across diverse packet-loss conditions and substantially better playback continuity under random and burst packet loss; even at a 70\% packet-loss ratio, TokJSNC maintains an average playback frame rate above 22~fps, whereas the strongest baseline remains below 12~fps. These results demonstrate the effectiveness of adaptive token transmission for robust real-time video streaming over lossy networks. TokJSNC provides a fundamental framework for video token communications.

\ifCLASSOPTIONcaptionsoff
\newpage
\fi

\bibliographystyle{IEEEtran}
\bibliography{Ref}
\end{document}